# Criteria on Utility Designing of Convex Optimization in FDMA Networks

Zheng SUN
School of Information Engineering, Beijing University of Posts and Telecommunications, Beijing, China
zhengs.bupt@gmail.com

*Abstract*—In this paper, we investigate the network utility maximization problem in FDMA systems. We summarize with a suite of criteria on designing utility functions so as to achieve the global optimization convex. After proposing the general form of the utility functions, we present examples of commonly used utility function forms that are consistent with the criteria proposed in this paper, which include the well-known proportional fairness function and the sigmoidal-like functions. In the second part of this paper, we use numerical results to demonstrate a case study based on the criteria mentioned above, which deals with the subcarrier scheduling problem with dynamic rate allocation in FDMA system.

I. INTRODUCTION

As the downlink access data rate increases, future wireless networks are expected to support various services with different quality of service (QoS) demands [1]. We can classify high speed services into two classes based on their delay tolerance. The QoS services are delay and rate sensitive, and require a certain access data rate. This type of application includes many high speed downlink data services that are widely studied over the last decades, such as *Video on Demand* (VOD) and packet-switched voice services. The other class corresponds to the best-effort services conducting more elastic applications such as file transfer and e-mail. This kind of services can adjust their data rate gradually and is often delay tolerant [2]. It is commonly believed that the concept of utility function, which maps the access data rate to the level of user satisfaction or QoS, is appropriate to characterize the elasticity of services.

In the past few years, utility-based radio resource management problems such as rate control and power allocation have been widely studied, and most of them dealt with the situations where the utility functions are either convex ones for which efficient theories and algorithms such as the Karush-Kuhn-Tucker (KKT) conditions exist, or specific nonconvex ones including the well-known sigmoidal-like function of which the dual problems are explored and solved by centralized or distributive algorithms. However, some research shows that the convex utility functions are appropriate only to model elastic services and do not capture the properties of services with strict QoS demands [2]. And the nonconvex utility maximization problem is significantly hard to be analyzed and solved, even by centralized computational methods. Particularly, nonconvexity makes a local optimum may not be a global optimum and there exists strictly positive dual gap. The standard distributive algorithms solving the dual problem may produce infeasible or suboptimal rate allocation, and the global maximization of nonconcave functions is an intrinsically difficult problem [3].

Due to these issues, in this paper, we would like to study the utility maximization problem in another way, by not specifying any particular forms of utility functions, even not assuming their convexity, but proposing sufficient and necessary conditions under which a utility function can guarantee leading the objective function convex. Although the content in this paper deals mainly with the network utility maximization problem in FDMA systems, the idea of exploring utility properties that achieves the global problem convex can be further studied in other scenarios, such as CDMA systems. In the numerical analysis, we will use the proposed criteria to demonstrate and solve an FDMA subcarrier scheduling problem with dynamic rate allocation. Various service types are considered, and we use different utility functions based on the criteria mentioned above to characterize the QoS demand properties of different services. For instance, for the VoIP and video streaming services, sigmoidal-like functions are appropriate to model their utilities, since decreasing the transmission data rate below a certain threshold would result in a significant drop in the QoS. And the utility of the best-effort service, which does not need a constant data rate support, would be more likely modeled using convex functions such as the logarithm function, since the more bits transmitted to the user, the more satisfied the user would be. However, it should be noticed that the sigmoidal-like function and the logarithm function are only two of many function forms that consistent with the criteria proposed in this paper. The main contribution of this paper is that we try to present a way which the researchers may follow to choose or design appropriate utility functions that could nicely model the characteristics of both real-time and best-effort services while guaranteeing that the optimization problem will always be convex.

The rest of the paper is organized as follows. In Section II, we present the system description and formulate the optimization problem. In Section III, we propose the criteria on utility designing and present an optimal power allocation algorithm. Based on the criteria mentioned, in Section IV we will use numerical results to demonstrate an FDMA subcarrier scheduling problem with dynamic rate allocation. And Section VI gives a summary and concludes the work.

This research is sponsored by Project 60772108 supported by National Natural Science Foundation of China and National Basic Research Program of China (973 Program), 2007CB310604

## II. GLOBAL NETWORK UTILITY MAXIMIZATION

In this section, we briefly describe the studied FDMA network scenario and formulate the general global network utility maximization problem. Consider a single-cell downlink Orthogonal FDMA system with N users. Inter-cell interference is not taken into consideration. The total system bandwidth $W$ is divided into $K$ subcarriers with bandwidth $\Delta f = W/K$. Let $H_{ik}$ denote the channel frequency response at subcarrier $k$ with user $i$, so the SNR of user $i$ at this subcarrier is expressed as $\eta_{ik} = \Gamma \cdot |H_{ik}|^2 p_{ik}/N_{ik} = \beta_{ik} p_k$, where $p_k$ is the transmit power allocated at subcarrier $k$, $\Gamma = -\ln(5\text{BER})/1.5$, and $N_{ik}$ is the noise power density. Let $r_{ik}$ denote the access data rate of user $i$ at subcarrier $k$, then $r_{ik} = \log(1 + \beta_{ik} p_k)$ (bit / s). The total achievable throughput of one user is defined as the sum of channel capacities of all its scheduled subcarriers. With a given subcarrier assignment, the total achievable throughput of user $i$ is given by $\sum_{k \in D_i} r_{ik}$. Consider the following problem of maximizing the network utilities for users:

$$\begin{aligned}
&\text{maximize} \sum_{i \in U} \sum_{k \in D_i} f_i(r_{ik}) \\
&\text{subject to} \sum_{i \in U} p_i \leq 1, p_i \geq 0, (\forall i \in U) \\
&\bigcup_{i \in U} D_i = D, D_j \bigcap D_i = \varnothing, (i \neq j, \forall i, j \in U)
\end{aligned} \quad (1)$$

where $f_i(x)$ is the utility on data rate of users $i$. Users conducting different services may have different utility function forms. $D$ and $U$ is the subcarrier set and the user set, and $D_i$ denotes the set of subcarriers assigned to user $i$.

## III. CRITERIA ON CONVEX UTILITY DESIGNING

Given that the problem in (1) is convex, it would be much convenient to utilize efficient theoretical or numerical algorithms, such as Interior-point method [6]. So in this setion, we will focus on determining in which conditions the problem (1) could be treated as a convex problem.

### A. Criteria on Utility Designing

***Theorem 1*: The problem (1) is a convex optimization problem, if and only if**

$$f_i(x) = \int e^x \left[ \int -t_i(x) e^{-x} dx + C_1 \right] dx + C_2, \quad \forall i \in U \quad (2)$$

**where $t_i(x) \geq 0$, $\forall x \geq 0$ and $C_1, C_2$ are real constants.**

*Proof*: Note that the constraint set and the domain of the objective function in (1) are both convex. This means if the objective function in (1) is concave, the problem achieves convex. Before showing Theorem 1, we first present following lemmas.

***Lemma 1*: The objective function in (1) is concave if and only if $f_i''(x) \leq f_i'(x)$, $\forall i \in U$.**

*Proof*: Notice that the maximization in problem (1) is equivalent to minimizing $-\sum_{i \in U} \sum_{k \in D_i} f_i(r_{ik})$, which can be equivalently rewritten as $y = -\sum_{k \in D} f_{i(k)}(r_{i(k)k})$, where $i(k)$ denotes the subcarrier scheduling result in which the subcarrier $k$ is assigned to user $i$. Then we have

$$\begin{cases} \dfrac{\partial^2 y}{\partial p_k^2} = -\dfrac{\beta_{i(k)k}^2}{\left(1 + \beta_{i(k)k} p_k\right)^2} \left( f_{i(k)}''(r_{i(k)k}) - f_{i(k)}'(r_{i(k)k}) \right) \\ \dfrac{\partial^2 y}{\partial p_i \partial p_j} = 0, \quad \text{if } i \neq j \end{cases} \quad (3)$$

From convex optimization theory we know that $y$ achieves convex if and only if the domain of $y$ is convex and its Hessian is positive semidefinite [6], i.e. $f_{i(k)}''(x) \leq f_{i(k)}'(x)$, $\forall x \geq 0$. ∎

Lemma 1 can be regarded as a guideline when designing utility functions for radio resource management in FDMA networks, which could be seen as one of the main contributions of this paper. It tells us *what kinds of* utility function forms guarantee (1) to be convex. Notice that in this paper we do not specify any particular form of $f_i(x)$, so that the conclusion drawn here could be used generally in designing utility functions in FDMA systems. More over, we conclude with the following lemma.

***Lemma 2*: $\forall i \in U$, if utility function $f_i(x)$ is a non-decreasing convex function or a non-increasing concave function on $\mathbf{R}^+$, the optimization problem in (1) is convex.**

*Proof*: $\forall i \in U$, from the result of Lemma 1, i.e. $f_i''(x) \leq f_i'(x)$, if $f_i''(x) \geq 0$, then $f_i(x)$ is convex. We have $f_i'(x) \geq f_i''(x) \geq 0$, which means $f_i(x)$ is non-decreasing. If $f_i'(x) \leq 0$, then $f_i(x)$ is non-increasing, then $f_i''(x) \leq f_i'(x) \leq 0$, which means $f_i(x)$ is concave. To sum up, when the utility function $f_i(x)$ is non-decreasingly convex, or non-increasingly concave, the optimization problem in (1) is convex. ∎

Now we continue the proof of Theorem 1 by presenting a more general utility designing criterion using the result in Lemma 1. Given that $f''(x) \leq f'(x)$ (user index $i$ is omitted for clarity), we introduce a function $t(x)$ that acts as a slack variable and satisfies $t(x) \geq 0$, $\forall x \geq 0$, then $f''(x)$ can be treated as $f'(x) - t(x)$. Denoting $f'(x)$ as $q(x)$, we have $q'(x) = q(x) - t(x)$. This equation is a first-order ordinary differential equation, which has a general solution form as $q(x) = e^{-\int P(x)dx} \left[ \int Q(x) e^{\int P(x)dx} dx + C_1 \right]$. With $P(x) = -1$ and $Q(x) = -t(x)$, finally we have $q(x) = e^x \left[ -\int t(x) e^{-x} dx + C_1 \right]$ and $f(x) = \int e^x \left[ -\int t(x) e^{-x} dx + C_1 \right] dx + C_2$. ∎

From Theorem 1 we can discover that the form of $f(x)$ is only determined by $t(x)$. The only necessary condition of $t(x)$ is that $t(x) \geq 0$, $\forall x \geq 0$, so there would be a huge freedom for us to choose and design utility functions. Let us specifically consider the term $\int t(x)e^{-x}dx$ in (2). Suppose $t(x)$ is nicely differentiable, we have

$$\int t(x)e^{-x}dx$$
$$= -t(x)e^{-x} - t'(x)e^{-x} \ldots - t^{(N)}(x)e^{-x} + \int t^{(N+1)}(x)e^{-x}dx \quad (4)$$
$$= \lim_{N \to \infty}\left\{-\sum_{i=0}^{N} t^{(i)}(x)e^{-x} + \int t^{(N+1)}(x)e^{-x}dx\right\}$$

where $t^{(N)}(x)$ denotes the $N$-order derivative of $t(x)$.

### B. Examples of Useful Utility Functions

In this part, we would like to demonstrate several useful function forms of $t(x)$ in designing utility function by using the result of above sections, and will highlight their explanations.

**Case 1:** When $t(x)$ is the finite-power of $x$, i.e. $t(x) = ax^K$, ($a \in R$, $K \in Z^+$, $K < +\infty$), $q(x) = e^x\left[-\int ax^K e^{-x}dx + C_1\right]$ and $f(x) = \int e^x\left[-\int ax^K e^{-x}dx + C_1\right]dx + C_2$. With the result of (4), we have $f(x) = \sum_{i=0}^{K}\left(aK! \cdot x^{K-i+1}/(K-i+1)!\right) + C_1 e^x + C_2$.

**Case 2:** When $t(x)$ is the finite-order polynomial of $x$, $t(x) = \sum_{n=0}^{K} a_n x^n$, $\left(a_n \in R, K \in Z^+, K < +\infty\right)$. This case is the linear weighted sum of Case 1. And the result is $f(x) = \sum_{n=0}^{K}\sum_{i=0}^{n} \frac{a_n n!}{(n-i+1)!} x^{n-i+1} + C_1 e^x + C_2$, which can be rewrite as $f(x) = \sum_{i=1}^{K+1}\left(\sum_{m=i-1}^{K} a_m \cdot m!\right)\frac{x^i}{i!} + C_1 e^x + C_2$. Note that $f(x)$ is a $K+1$-order polynomial of $x$, which can be used as a polynomial fit of the empirical data of the utility function between user experience and data rate. Specifically, given that a polynomial fit for the empirical data is $\tilde{f}(x) = \sum_{i=0}^{N} \tilde{a}_i x^i$. Compare it with $f(x)$, we have $C_1 = 0$, $C_2 = \tilde{a}_0$ and

$$\sum_{i=m-1}^{N} a_i \cdot i! = \tilde{a}_m \cdot m!, \forall m \in \{1 \ldots N\}. \quad (5)$$

Equations (5) formulate an $N$-variable matrix equation, which can be efficiently solved as $a_i = \tilde{a}_i/i!$ (if $i = N$) and $a_i = (\tilde{a}_{i+1} - \tilde{a}_i)/i!$ (if $i < N$).

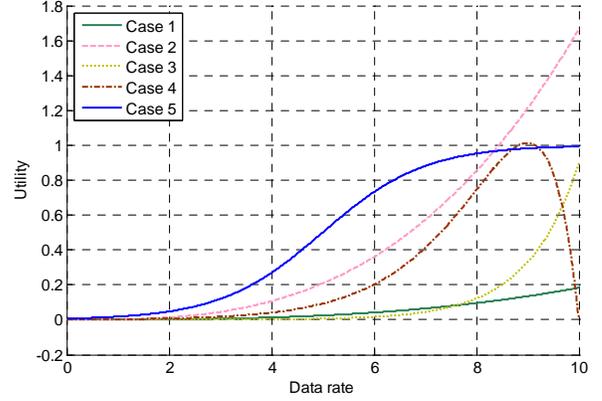

Fig. 1. Some examples of various possible utility function forms that guarantee the optimization problem convex based on the criteria of Theorem 1. Notice that the convexity of any specific utility function is not presumed.

**Case 3:** $t(x)$ is the exponential function of $x$: $t(x) = e^{ax}$, $a \in R$. When $a \neq 1$, $f(x) = \frac{e^{ax}}{a(1-a)} + C_1 e^x + C_2$ and when $a = 1$, $f(x) = -xe^x + (C_1 + 1)e^x$.

**Case 4:** Proportional fairness: if

$$t(x) = \frac{C_0 C_1}{C_1 x + C_2} + \frac{C_0 C_1^2}{(C_1 x + C_2)^2}, \quad (6)$$

we have $f(x) = C_0 \log(C_1 x + C_2) + C_3 + C_4 e^x$. Notice that if $C_3 = 0$ and $C_4 = 0$, $f(x)$ is the well-known proportional fairness utility function of the data rate, which has been extensively studied [5], [7].

**Case 5:** Sigmoidal-like function: Specifically, let us consider a special case when $t(x) = \frac{2e^{2x+x_0}}{(e^{x_0} + e^x)^3}$, therefore $f(x) = \left(1 + e^{-x+x_0}\right)^{-1} + C_1 e^x + C_2$. Given that $C_1 = 0$ and $C_2 = 0$, we can see that $f(x)$ becomes the well-known sigmoidal-like function having an inflection point at $x_0$, which has been considered to have impressive second-order differential property to model the elasticity of delay and rate sensitive services such as video streaming and VoIP [2].

From above, the criteria proposed in this paper can be regarded as a summary of several commonly used utility designing strategies in communication system, which have been studied and researched for many years. The most important contribution of this paper is that we present a way which the researchers may follow to choose or design appropriate utility functions that nicely model the characteristics of both real-time and best-effort services while guaranteeing that the optimization problem will always be convex.

### C. Optimal Power Allocation

In this section, we will propose an optimal power allocation algorithm for the FDMA system which has been built based on the criteria mentioned above. Suppose that $\forall i \in U$, $f_i(x)$'s satisfy Theorem 1, i.e. the objective problem described in (1) is convex, therefore the Lagrangian of (1) is

$$L(\lambda, \upsilon) = -\sum_{k \in D} f_{i(k)}\left(r_{i(k)k}\right) + \sum_{k \in D} \lambda_k p_k + \upsilon\left(\sum_{k \in D} p_k - 1\right). \quad (7)$$

From KKT conditions, we have

$$\begin{cases} \upsilon\left(\sum_{k \in D} p_k - 1\right) = 0 \\ \lambda_k p_k = 0, \forall k \in D \\ \nabla y + \sum_{k \in D} \lambda_k \nabla p_k + \upsilon \nabla\left(\sum_{k \in D} p_k - 1\right) = 0 \end{cases} \quad (8)$$

From (8), consider $\lambda_k$'s as slack variables, we rewrite KKT conditions as follows.

$$\begin{cases} \dfrac{\beta_{i(k)k} f'_{i(k)}\left(\ln\left(1 + \beta_{i(k)k} p_k\right)\right)}{1 + \beta_{i(k)k} p_k} \leq \upsilon \\ \left(\upsilon - \dfrac{\beta_{i(k)k} f'_{i(k)}\left(\ln\left(1 + \beta_{i(k)k} p_k\right)\right)}{1 + \beta_{i(k)k} p_k}\right)(-p_k) = 0 \end{cases}, \forall k \in D. \quad (9)$$

The problem in (9) can be readily solved using centralized and distributed iterative algorithms by updating the value of $\upsilon$ with specific step sizes [8].

If $f_i(x) = x$, $\forall i \in U$, the problem in (9) will reduce to the traditional waterfilling solution as

$$p_k = \begin{cases} \dfrac{1}{\upsilon} - \dfrac{1}{\beta_{i(k)k}} &, \text{if } \upsilon < \beta_{i(k)k} \\ 0 &, \text{otherwise} \end{cases}, \forall k \in D, \quad (10)$$

which can be conveniently solved by using iterative or exact algorithms shown in [9].

## IV. NUMERICAL ANALYSIS

In this section, the proposed criteria on utility designing are demonstrated by an FDMA subcarrier scheduling problem with dynamic rate allocation. We divide the whole problem into two steps similar as in [10]. For the first step, the subcarrier scheduling is done based on the channel conditions without accounting the power level assigned to each subcarrier. When the subcarrier assignment is over, the multiuser OFDM system can be considered as a FDMA system with dynamic subcarrier allocation [10]. So in the second step, we will use the conclusion drawn in Section III.C to accomplish the power allocation.

During the analysis, the total 1.024 MHz bandwidth is divided into 256 subcarriers. The cell radius is set to 1 km and

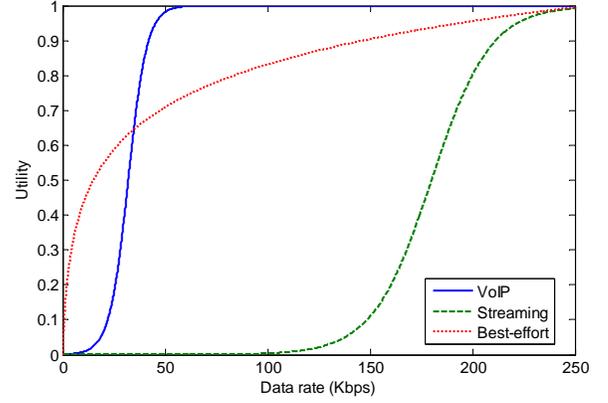
Fig. 2. Utility functions used in numerical analysis section.

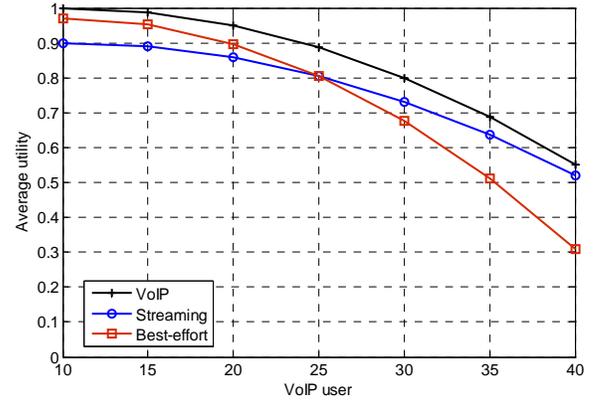
Fig. 3. Average utility when increasing VoIP user number.

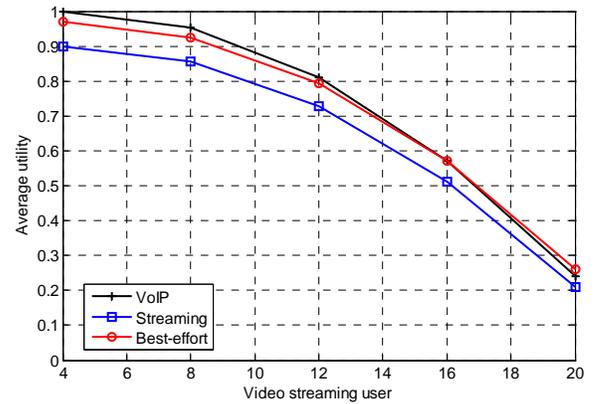
Fig. 4. Average utility when increasing video streaming user number.

the path loss factor $PL(d) = 38.4 + 20\log_{10} d$ [dB], where $d$ (m) is the distance between users and the base station. Shadowing is assumed to be lognormally distributed with mean 0 dB and standard deviation 8 dB. Every user moves at an average speed of 20 m/s. The standard deviation of user speed is 2.24 m/s. Each user is dedicated to one session of a specific service type. The scheduling performs every 0.125 ms. The transmit power from the base station is fixed to 43dBm, and the thermal noise power is -108 dBm. The achievable coding rates are {1/2, 2/3, 3/4, 7/8}. The selective modulation schemes are QPSK, 16QAM, 32QAM, and 64QAM.

We consider three services: VoIP, video streaming and best-effort traffic. The VoIP traffic is generated according to the ON/OFF model in [13]. The average durations of ON and OFF periods are 1.0 s and 1.5 s respectively. We assume within each ON interval, the voice data rate is 32 Kbps, and the lifetime of a packet is 80 ms. The video streaming traffic is according to the model in [14]. The duration of each state is exponentially distributed with mean 160 ms. The data rate of each state is in a truncated exponential distribution where the data rate range is from 64 to 256 with an average value of 180 Kbps, and the maximum packet delay is 1s. In order to simulate the maximum performance of the best-effort traffic, we apply a full-buffer model, so that the maximum throughput for the best effort services can be obtained.

Without loss of generality, we let VoIP users have a sigmoidal utility function with inflection point at 32 Kbps (corresponding to Case 5 in Section III. B), video streaming users a sigmoidal utility function with inflection point at 180 Kbps, and best-effort user a logarithm utility function (corresponding to Case 4 in Section III. B). During the simulation, we normalize the utility function such that $\forall i \in U$, $f_i(0) = 0$ and $f_i(M_i) = 1$, where $M_i$ is the maximum transmission data rate of user $i$ (note that it is not necessary to normalize the utility function). The utility functions of services are plotted in Figure 2.

Firstly we fix the number of video streaming and best-effort users to be 4 and 20 respectively, and increase the number of VoIP users from 10 to 40. It can be seen from Figure 3 that as the number of VoIP users increases, the average utility of every service drops down since the radio resource gets more and more scarce. The similar situation appears in Figure 4, when we increase the number of video streaming users from 4 to 20, and fix the number of VoIP users and BE users to be 10 and 20 respectively. Because the video streaming sessions have much higher access data rates, i.e. 180 Kbps for average, the utility values drop down with a higher speed than that of Figure 1.

## V. CONCLUSIONS

In this paper, we have studied the network utility maximization problem in FDMA networks and summarize with a suite of criteria on designing utility functions so as to make the global optimization convex. After we proposed the general form of the utility functions in Theorem 1, five cases of commonly used utility forms have been presented and their usages and explanations have been discussed. By showing that the famous sigmoidal-like functions and the proportional fairness function are consistent with the criteria proposed in this paper, we conclude that this paper may be seen as a summary of several widely used utility designing strategies studied in recent years.

In the second part of this paper, after introducing the optimal power allocation algorithm, we used numerical analysis to demonstrate a case study based on the criteria, which deals with the subcarrier scheduling problem with dynamic rate allocation in FDMA system. Numerical results show that the criteria and the optimal power allocation algorithm proposed in this paper can be well utilized in FDMA systems.

REFERENCES

[1] Y.Cao and V.O.K.Li, "Scheduling algorithms in broad-band wireless networks", *IEEE Proc. the IEEE*, vol. 89, pp. 76-87, Jan. 2001.

[2] J. W. Lee, R. R. Mazumdar and N. B. Shroff, "Non-convex optimization and rate control for multi-class servises in the internet", *IEEE/ACM Trans. on Networking*, vol. 12, No. 4, Augest 2005.

[3] M. Fazel and M. Chiang, "Network utility maximization with nonconcave utilities using sum-of-squares method", *Proc. of 44th IEEE Conf. on Decision and Control*, Seville, Spain, December 12-15, 2005.

[4] O. Shin, K. B. Lee, "Packet Scheduling over a Shared Wireless Link for Heterogeneous Classes of Traffic," *IEEE ICC*, vol. 1, pp. 58-61, 20-24 June 2004.

[5] T. D. Nguyen and Y. Han, "A proportional fairness algorithm with QoS provision in downlink OFDMA system", *IEEE Comm. Letter*, vol. 10, No. 11, November 2006.

[6] S. Boyd and L. Vandenberghe, *Convex Optimization,* Cambridge University Press, 2004.

[7] H. Seo and B. G. Lee, "Proportional-fair power allocation with CDF-based scheduling for fair and efficient multiuser OFDM systems", *IEEE Trans. on Wireless Comm.*, vol. 5, No. 5, May 2006.

[8] B. Johansson, P. Soldati and M. Johansson, "Mathematical decomposition techniques for distributed cross-layer optimization of data networks", *IEEE J. Select. Areas Commun.*, vol. 24, No. 8, August 2006.

[9] D. P. Palomar and J. R. Fonollosa, "Practical algorithms for a family of waterfilling solutions", *IEEE Trans. on Signal Processing*, vol. 53, No. 2, February 2005.

[10] J. Jang and K. B. Lee, "Transmit power adaptation for multiuser OFDM systems", *IEEE J. Select. Areas Commun.*, vol. 21, No. 2, February 2003.

[11] R. Knopp and P. Humblet, "Information capacity and power control in single cell multiuser communications", in *Proc. IEEE Int. Conf. Commun.*, Seattle, WA, June 1995, pp. 331-335.

[12] A. Jalali, R. Padovani, and R. Pankaj, "Data throughput of CDMA-HDR a high efficiency-high data rate personal communication wireless system", in *Proc. IEEE. Veh. Technol. Conf. Spring*, Tokyo, Japan, May 2000, pp. 1854-1858.

[13] A. Sampath and J. M. Holtzman, "Access control of data in integrated voice/data CDMA systems: benefits and tradeoffs," *IEEE J. Select. Areas Commun.*, Oct. 1997, pp. 1511-1526.

[14] Decina, M. and Toniatti, T., "Bandwidth allocation and selective discarding for a variable bit rate video and bursty data calls in ATM networks," *Int'l J. Digital Analog Commun. Systems*, vol. 5, pp. 85–96, Apr.-June 1992.

[15] S. Shakkottai and A. L. Stolyar, "Scheduling algorithms for a mixture of real-time and non-real-time data in HDR", *Bell Labs Technical Report*, Oct. 2000.